# Environmental Effects in Mechanical Properties of Few-layer Black Phosphorus

Miriam Moreno-Moreno[1], Guillermo Lopez-Polin[1], Andres Castellanos-Gomez[2], Cristina Gomez-Navarro[1,3]* and Julio Gomez-Herrero[1,3].

[1]Departamento de Física de la Materia Condensada, Universidad Autónoma de Madrid, 28049, Madrid, Spain.

[2]Instituto Madrileño Estudios Avanzados IMDEA Nanociencia, Madrid 28049, Spain.

[3]Centro de Investigación de Física de la Materia Condensada (IFIMAC), Universidad Autónoma de Madrid, 28049, Madrid.

**Abstract**

*We report on the mechanical properties of few-layer black phosphorus (BP) nanosheets, in high vacuum and as a function of time of exposure to atmospheric conditions [1]. BP flakes with thicknesses ranging from 4 to 30 nm suspended over circular holes are characterized by nanoindentations using an atomic force microscope tip. From measurements in high vacuum an elastic modulus of 46±10 GPa and breaking strength of 2.4±1 GPa are estimated. Both magnitudes are independent of the thickness of the flakes. Our results show that the exposure to air has substantial influence in the mechanical response of flakes thinner than 6 nm but small effects on thicker flakes.*

The recent isolation of atomically thin materials from bulk layered crystals by mechanical or chemical exfoliation [2, 3] holds promise to revolutionize the field of flexible electronics. Graphene is still by far the most studied two-dimensional crystal, however, other 2D crystals have recently gained considerable interest as they present similar mechanical performance while their electronic properties are complementary to those of graphene [4]. The lack of bandgap in graphene, for example, has motivated a surge of works on semiconducting 2D materials [5-8] which can also have a strong impact in future flexible electronics. Indeed, transition metal dichalcogenides (TMDCs) (Mo- and W- based dichalcogenides, mainly) have been recently used to fabricate flexible field effect transistors and photodetectors due to their high flexibility and high breaking strength [9-12].

TMDCs, however, present relatively slow photoresponse and, because of the large bandgap of Mo- and W-based compounds, they are only suited for applications in a limited part of the visible range of the electromagnetic spectrum. A material with a direct and narrow bandgap together with fast photoresponse is needed to extend the detection range accessible with 2D materials. Recently, black phosphorus thin flakes have been isolated by mechanical exfoliation of bulk synthetic crystals [13-17]. Bulk black phosphorus is a semiconductor with a direct bandgap of 0.35 eV with relatively high charge carrier mobilities in the order of 1000 cm$^2$/Vs. Both values change with the number of layers, reaching more than 2 eV and around 1 cm$^2$/Vs respectively [18, 19] in its single layer form. Remarkably, strain provides another way of tuning electrical and optical properties [20-22]. A distinctive feature of BP is the in-plane anisotropy in its transport, optical and mechanical properties due to its puckered structure [23-25]. As examples (i) the in-plane anisotropy is around 64% for the electrical conductance (which is in the order of 10 μS for few-layer BP) [25] and (ii) the prominent electronic transport direction (armchair) is orthogonal to the prominent heat transport direction (zig-zag) [26].

A significant issue in BP is its environmental instability: BP is unstable in ambient conditions in its single- and few-layer form [27]. Specifically, the presence of moisture and oxygen in air leads to the degradation of the material due to the formation of oxidized phosphorus species [28]. While some studies have already addressed environmental effects in the electronic properties of BP [27, 29-32], little is still known about its effect on the mechanical properties.

Here we investigate the role of air ambient exposition in the mechanical properties of BP. Atomic force microscopy indentation experiments carried out in high vacuum conditions (HV, ~10$^{-6}$ mbar) yield elastic modulus of $E_{3D}$=46±10 GPa for flake thicknesses ranging between 4 and 30 nm. Due to the geometry of our experiments (indentations at the center of BP circular membranes), the obtained $E_{3D}$ is an average of the elastic modulus in both in-plane directions (zig-zag and armchair). Considering this, the measured value of $E_{3D}$ is in good agreement with previous measurements in atmospheric conditions [25] and with theoretical predictions [33-36]. Our results also show that the exposure to air has a small effect on flakes thicker than 7 nm, even for time periods as long as 200 hours, but substantial influence in the mechanical response of flakes thinner than 6 nm (corresponding to 10-12 layers).

For this study BP drumheads were prepared by mechanical exfoliation of bulk black phosphorus (Smart Elements) on SiO$_2$ (300 nm)/Si substrates with predefined circular wells with diameters ranging from 0.5 to 3 µm (figure 1(a, c) and Supplementary Information 1). The mechanical properties of the membranes were tested by indenting with an atomic force microscopy (AFM) tip at the center of the suspended area as sketched in figure 1(b) (details about the AFM probe can be found in Supplementary Information 2).

During the sample preparation procedure BP samples were not exposed to atmospheric conditions for more than 2 hours (below this time no substantial degradation is appreciated [27]). Samples were kept and initially measured in a HV chamber. Subsequent to a complete characterization of their mechanical properties in HV conditions, a set of measurements in each membrane was performed while they were exposed to environment to assess the influence of ambient conditions in few-layer BP mechanical properties.

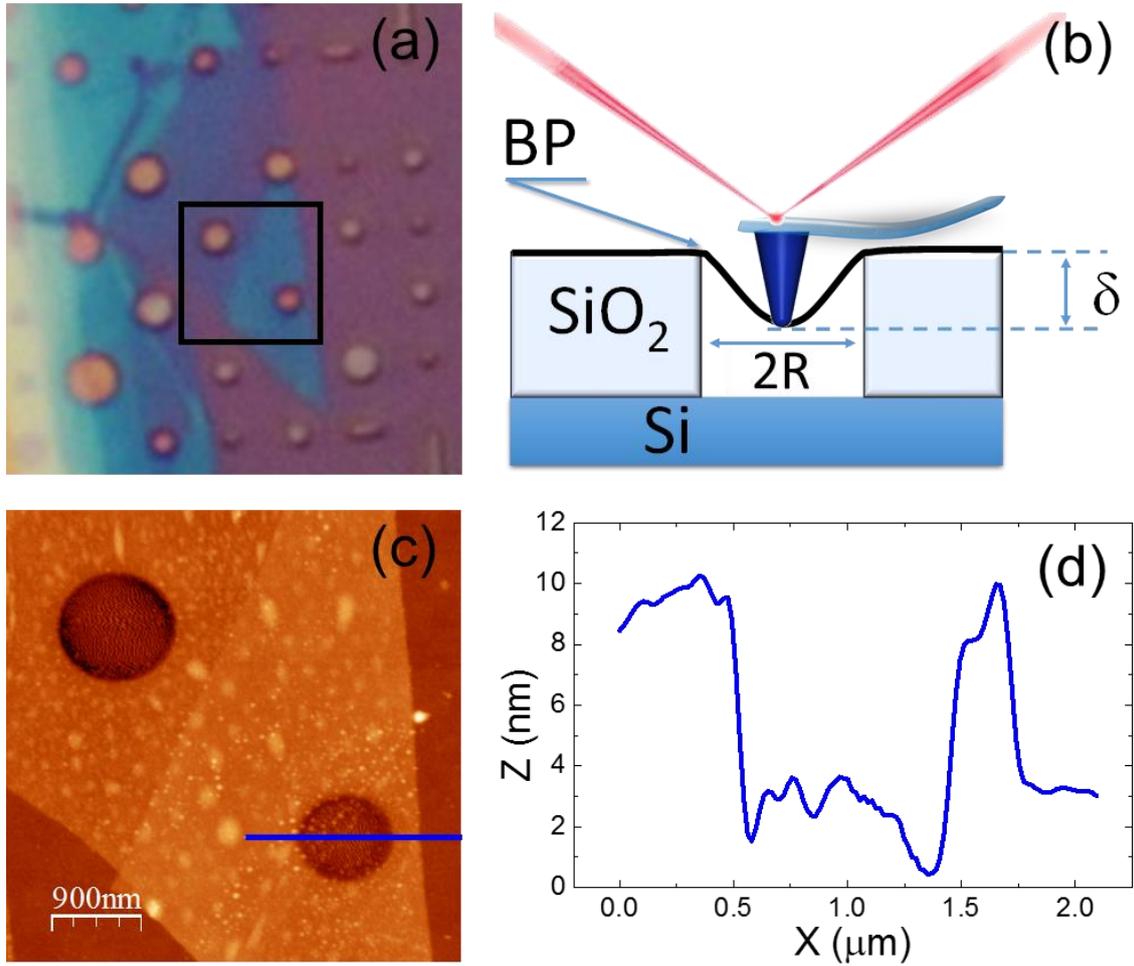

**Figure 1.** Sample geometry and set up of the nanoindentation test experiments. (a) Optical microscopy image of representatives flakes deposited on an array of circular wells. (b) Schematic diagram of nanoindentations performed on the BP drumheads. (c) AFM topography image (zoom in of (a)) of a BP flake covering two wells. (d) Topographic profile along the blue line in (c).

AFM indentation experiments on freely suspended BP can be modeled as clamped circular drumheads with central point loading (Supplementary Information 3). The force $F$ versus indentation $\delta$ curves can be approximated by [10]

$$F = \left[\frac{4\pi E_{3D}}{3(1-v^2)} \cdot \left(\frac{t^3}{R^2}\right)\right] \cdot \delta + \pi\sigma_0 \cdot \delta + \left(\frac{q^3 E_{3D} t}{R^2}\right) \cdot \delta^3 \qquad [1]$$

where $E_{3D}$ and $v$ are the elastic modulus and the Poisson's ratio respectively (the last one taken as 0.45, as the average of the Poisson's ratio in the [100] and in the [010] direction, from Ref. [35]), $t$ is the thickness of the nanosheet, $R$ its radius, $\sigma_0$ the pre-tension and $q = 1/(1.05-0.15v-0.16v^2)$ is a dimensionless parameter close to 1. The first term in the equation is associated with the bending rigidity and is negligible for few layered flakes. The second term, also linear in $\delta$, accounts for the effect of small pre-stress accumulated in the membrane during the fabrication method and varies quite randomly from drumhead to drumhead. Finally, the term in $\delta^3$ dominates at large indentations and is governed by the intrinsic elastic modulus of the sheets.

The thickness and deflection dependences reflected in Equation [1] provide two complementary methods for determining $E_{3D}$. In the first method the experimental data of each indentation curve are fitted to a polynomial with terms in $\delta$ and $\delta^3$ and the elastic modulus is extracted from the cubic coefficient. It requires large indentations but the value of $E_{3D}$ can be obtained with data of exclusively one drumhead. This method automatically separates the linear coefficients from the elastic modulus, however it is not easy to apply to thick flakes where the large indentation regime is difficult to achieve. Therefore, by using this method, the elastic modulus is directly obtained from each indentation curve, however, the bending rigidity and pre-tension are not separately accessible since both contribute to the linear coefficient. The second method consists in measuring a set of flakes with different thicknesses and/or radii in the low indentation regime, where the term in $\delta^3$ is negligible. Therefore the $\partial F/\partial\delta$ (the spring constant of each flake, $k_{flake}$) scales as $t^3/R^2$ [10]. Consequently, the elastic modulus is inferred by linear fitting of $k_{flake}$ versus $t^3/R^2$ for several drumheads, according to this expression:

$$k_{flake} = \left.\frac{\partial F}{\partial \delta}\right|_{\delta\approx 0} = \frac{4\pi E_{3D}}{3(1-\nu^2)} \cdot \left(\frac{t^3}{R^2}\right) + \pi\sigma_0 \qquad [2]$$

This second method is valid while the pre-tensions of all the drumheads are similar. Under that assumption, a linear fit would provide information about the Young's modulus (from the slope) and the *average* pre-tension of them (from the intercept).

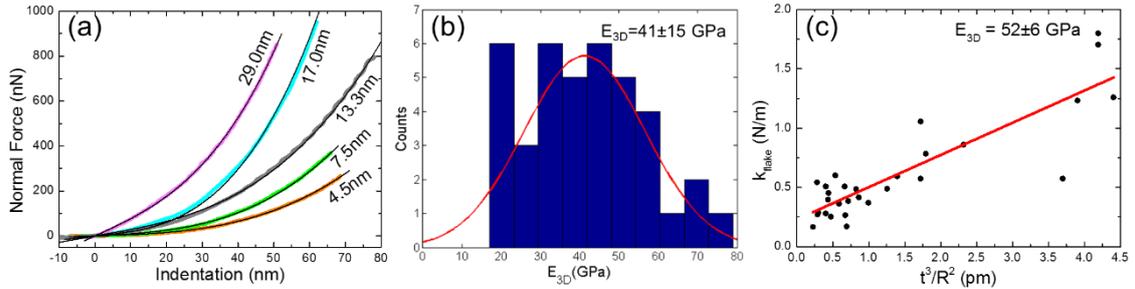

**Figure 2.** Indentation curves and $E_{3D}$ of BP drumheads in high vacuum. (a) Force versus indentation ($F(\delta)$) curves (coloured lines) performed in HV in five BP drumheads with different thickness (4.5, 7.5, 13.3, 17 and 29 nm) and their cubic polynomial fit to Eq. (1) (thin black lines). (b) Histogram of the $E_{3D}$ values obtained from the cubic polynomial fit of $F(\delta)$ curves performed in 39 BP drumheads, following the first method described in the main text. The fit of the data of the histogram to a normal distribution yields $E_{3D}$= 41±15 GPa. (c) Elastic constant versus $t^3/R^2$ measured for 29 BP drumheads, and their linear fit to the Expression [2] (red solid line). This method yields $E_{3D}$= 52±6 GPa.

Figure 2 summarizes our main findings in high vacuum conditions. Figure 2(a) portrays up to 5 representative indentation curves acquired on flakes with thicknesses ranging from 4.5 up to 29 nm. Figure 2(b) displays a histogram of the values obtained for the elastic modulus of 39 different flakes as obtained employing the first method aforementioned. The average value for $E_{3D}$ is 41±15 GPa, without any appreciable tendency with flake thickness. The pre-stress obtained from the linear term are in the range of 0.05 up to 0.3 N/m (obtained exclusively from flakes thinner than 5 nm where the bending rigidity is much lower that the pre-stress). Figure 2(c) depicts a set of experimental data points obtained in the low indentation regime, and it plots $k_{flake}$ vs. $t^3/R^2$, following the second method described above. The obtained elastic modulus value is 52±6 GPa, within the error of the elastic modulus obtained in figure 2(b), and an average pre-tension of 0.07±0.02 N/m.

In order to analyze the influence of random pretensions in the flakes (which would lead to different intercepts in the linear fit of figure 2(c)), we have numerically simulated $F(\delta)$ curves with random pre-tensions in the low indentation regime (see Supplementary Information 4). Here we observe that the dispersion of the simulated data is similar to that of our experiments. This suggests that the variation of pre-tension among nanosheets is the main source of noise in the experimental data.

Summarizing, both methodologies yield similar results of the elastic modulus of few layered BP, being 46±10 GPa the weighted average of both values (taking into account the number of drumheads analyzed with each method).

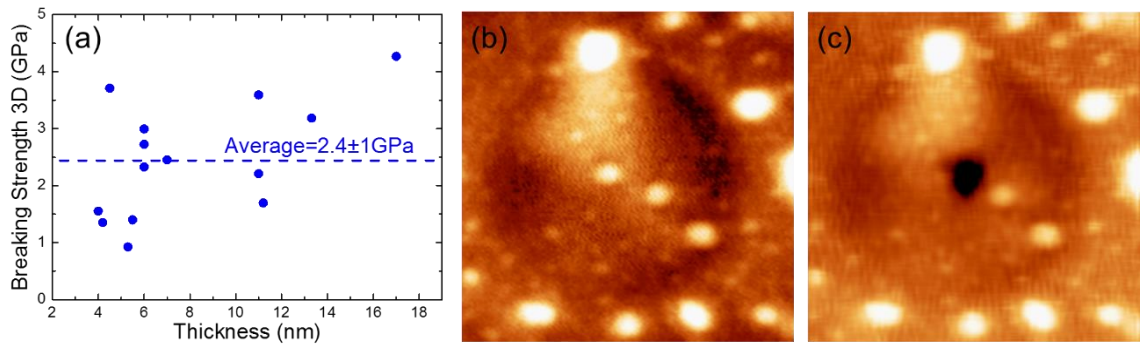

**Figure 3**. (a) Breaking strength of 14 BP drumheads versus the BP thickness measured in high vacuum (having an average breaking strength of 2.4±1 GPa). (b) AFM image of a BP drumhead in high vacuum before its breaking. (c) AFM image of the same BP drumhead as in (b) after its rupture, also in high vacuum. The size of the AFM images is 1.5x1.5µm$^2$.

Towards measuring the breaking strength ($\sigma_{max}$) of BP, some of the drumheads were indented until rupture. In order to estimate the strength we assume that it can be expressed as [37] $\sigma_{max}=(F_{break} E_{3D}/4\pi R_{tip}t)^{1/2}$, where $F_{break}$ is the rupture force and $R_{tip}$ the tip radius. Figure 3(a) depicts a chart of the breaking strength vs. thickness for 14 BP drumheads measured in high vacuum. The plot does not show any clear correlation between these two magnitudes. The average breaking strength is found to be 2.4±1 GPa. This value compares quite well with previous measurements in atmospheric conditions

[25]. Figure 3(b) and (c) displays a BP membrane before and after breaking. As it can be readily seen, the crack is confined around the tip-sample contact point in contrast with other 2D materials such as $MoS_2$ or graphene [9, 38].

Subsequent to mechanical characterization in HV conditions, measurements in ambient atmosphere were performed. Once the samples were exposed to ambient conditions, measurements separated by intervals of few hours were carried out in each membrane. The relative humidity was ≈47% and the temperature ≈22°C. This scheme allows monitoring the evolution of $E_{3D}$ with the exposure time to ambient conditions. For the analysis of the nanoindentation measurements at atmospheric exposure, the first method described above was used. The results obtained from these experiments are depicted in figure 4(a), which shows the evolution of $E_{3D}$ as a function of the exposure time to atmosphere.

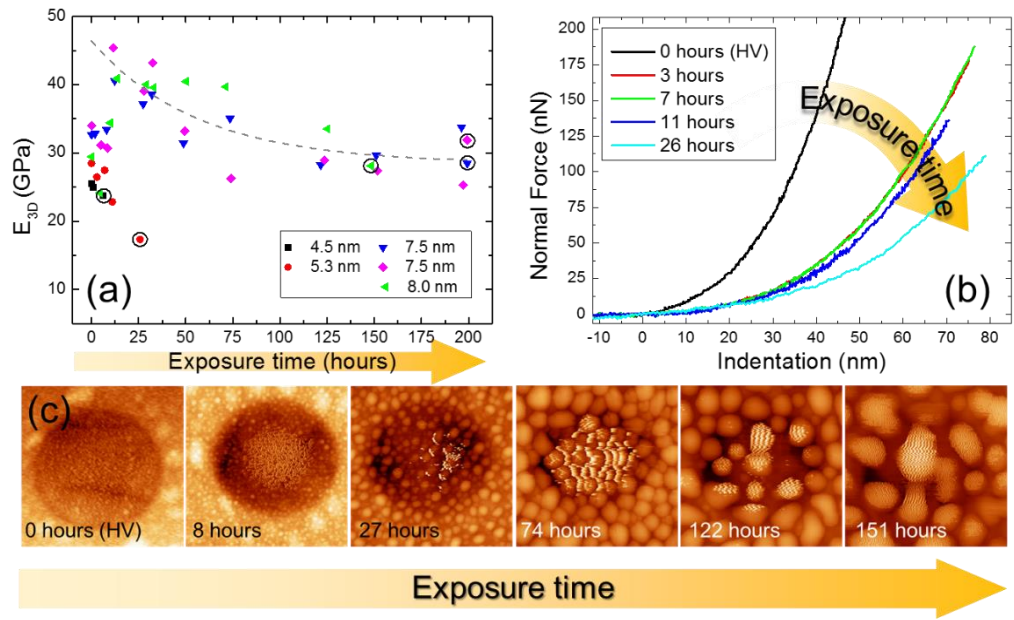

**Figure 4.** Evolution of the mechanical properties and topography of the BP drumheads under ambient conditions. (a) $E_{3D}$ of five drumheads versus the time of exposure to ambient conditions and after heating in high vacuum. Data points sharing color and shape correspond to the same BP drumhead, whose thicknesses range from 4.5 to 8 nm. The data points plotted inside a black circumference correspond to the last set of indentations performed on each drumhead, after that set the drumhead broke. The data plotted at 0 hours correspond to measurements in HV (before exposure), and the two last data plotted at almost 200 hours (and inside a black circumference) correspond to data acquired of two drumheads after the whole exposure (nearly 200 hours) and the subsequent heating of the sample in HV. The dashed line is the fitting of data of flakes thicker than 7 nm to a model proposed in Supplementary Information 6. (b) $F(\delta)$ curves for the 5.3 nm thick BP drumhead (plotted with red circles in (a)) in HV and after 3, 7, 11 and 26 hours of exposure to ambient conditions. In the last indentation (after 26 hours) the drumhead broke. (c) Set of AFM images showing the evolution of the topography of a 7.5 nm thick BP drumhead under the exposure to ambient conditions (from 0 hours of exposure, i.e. HV conditions, up to 151 hours of exposure), the size of the images is 1.5x1.5μm$^2$.

According to our measurements BP flakes with thickness above 7 nm show a minor reduction of its elastic modulus upon ambient conditions exposure suggesting the growth of a passivation layer. This is readily observed in the data plotted with blue, pink and green symbols in figure 4(a). We have considered a simple passivation process (see Supplementary Information 6). According to this model, an exponential decay fitting of these three sets of data is depicted by a dashed line. On the other hand, flakes with thickness below 6 nm show a clear tendency of decreasing elastic modulus when exposed to air (black squares and red dots in figure 4(a)), reaching a decrease of a factor of almost two at 24 hours of exposure to ambient. This fast decreasing in $E_{3D}$ for drumheads thinner than 6 nm can be clearly appreciated in figure 4(b), which shows several indentation curves acquired at increasing exposure time in the 5.3 nm thick flake of figure 4(a). It is plain to see that the elastic modulus decreases with time in this case. Indeed, the 4.5 nm thick flake spontaneously broke after the first 10 hours of exposure. Furthermore, all the flakes studied under ambient conditions, broke at forces well below forces that were previously supported.

As previously reported [27], our AFM topographic images of BP drumheads showed significant changes as a function of time of ambient exposure. Figure 4(c) displays AFM images showing the evolution of the topography of a representative BP flake with exposure time. As a consequence of water absorption we observe a random population of protrusions increasingly covering the surface. It should be noticed the difference between bubbles, that were present immediately after sample preparation, and protrusions originated by moisture. It is also worth mentioning that the fitting errors of the indentation curves also increase with time, most probably due to the presence of these protrusions, which slightly alter the shape of these curves (see Supplementary Information 3).

In brief: the main conclusions derived from our latter experiments are three: (i) although exposure to air leads to substantial changes in the topography of the few-layer BP, the elastic modulus of flakes thicker than 7 nm is slightly reduced, (ii) BP flakes thinner than 6 nm, experience a significant decrease of their $E_{3D}$ upon exposure to ambient conditions, and (iii) for both thickness ranges the rupture forces seem to decrease with the ambient degradation.

The elastic modulus of the flakes upon degradation yields the magnitude $E_{3D} \cdot t = E_{2D}$. For the results plotted in figure 4, the value of thickness, $t$, used to derive $E_{3D}$ is the height of the flake measured by AFM under high vacuum conditions. Upon exposure to atmosphere this magnitude becomes experimentally inaccessible for some time, i.e. the topography of the flakes reflects a growing height due to water absorption on the flakes that does not correspond to their actual thickness. Hence, the decrease of elastic modulus observed in our measurements might be ascribed to two different physical origins: (i) an inherent decrease of the $E_{3D}$ of the material and/or (ii) a decrease of the thickness of the membrane. In order to investigate these effects the samples were annealed in high vacuum at 230 °C during 15 hours. Following previous works, under these conditions a removal of adsorbed water occurs. This allows measuring the real thickness of the flakes. According to our AFM images before and after annealing (see Supplementary Information 5) the change in thickness is negligible, hence suggesting that a decrease of the intrinsic elastic constants is taking place during exposure to atmospheric conditions.

This decrease in $E_{3D}$ could be attributed to the passivation of the outer layers of the BP flakes leading to phosphorus oxide layers with a lower elastic modulus than that of pristine BP. Therefore, we consider the previously introduced passivation model (see Supplementary Information 6), in which the passivation layer depth increases at a rate that exponentially decreases with time, and the total thickness of the drumhead is constant in the whole process (as figure S7 in Supplementary Information 5 suggests). In order to draw conclusions from this model, determination of flake's thickness is paramount. As the real thickness of the flakes is now available, it is possible to complete the passivation model, which yields a maximum passivation depth of 9±5 nm and a passivation characteristic time of 60±20 hours. The reduction factor of elastic modulus of the phosphorene oxide with respect to the one of pristine phosphorene is reported to be around 0.66 [39], leading to an elastic modulus of 30±7 GPa for fully oxidized few-layer BP (46±10 GPa was taken as elastic modulus of pristine few-layer BP).

According to our model, the maximum of the elastic modulus should occur at 0 hours of exposure. However, the experimental results for flakes thicker than 7 nm show an increase of $E_{3D}$ for low exposure times. This behavior was theoretically predicted for phosphorene oxide in Ref. [39] and it is rationalized in terms of small relaxations of phosphorene due to chemisorbed oxygen atoms.

The presented results are in good agreement with those published by Tao *et al.* [25] of mechanical characterization in air of thick (> 14 nm) flakes but in clear disagreement with a more recent publication by Wang *et al.* [40].

In conclusion, in this work high vacuum measurements of BP mechanical properties are reported. Indentation experiments yield elastic modulus of 46±10 GPa and breaking strength of 2.4±1 GPa. The elastic modulus barely decreases in atmospheric conditions for thick flakes but we found a clear decreasing tendency for the thinnest flakes measured in our experiments. This is attributed to a self-passivation process that saturates with time.

**Acknowledgments:**
Financial support was received from MAT2013-46753-C2-2-P that includes Miriam Moreno-Moreno's FPI fellowship, CSD2010-0024, "María de Maeztu" Programme for Units of Excellence in R&D (MDM-2014-0377). AC-G also acknowledges financial support from the BBVA Foundation through the fellowship "I Convocatoria de Ayudas Fundacion BBVA a Investigadores, Innovadores y Creadores Culturales" ("Semiconductores ultradelgados: hacia la optoelectronica flexible"), from the

MINECO (Ramón y Cajal 2014 program, RYC-2014-01406) and from the MICINN (MAT2014-58399-JIN). We also acknowledge MAD2D-CM, S2013/MIT-3007.

# SUPPLEMENTARY INFORMATION

## Environmental Effects in Mechanical Properties of Few-layer Black Phosphorus

Miriam Moreno-Moreno[1], Guillermo Lopez-Polin[1], Andres Castellanos-Gomez[2], Cristina Gomez-Navarro[1,3]* and Julio Gomez-Herrero[1,3].

[1]Departamento de Física de la Materia Condensada, Universidad Autónoma de Madrid, 28049, Madrid, Spain.

[2]Instituto Madrileño Estudios Avanzados IMDEA Nanociencia, Madrid 28049, Spain.

[3]Centro de Investigación de Física de la Materia Condensada (IFIMAC), Universidad Autónoma de Madrid, 28049, Madrid.

### SI1. SAMPLE PREPARATION

Suspended black phosphorus flakes were prepared by deterministic placement over substrates of Si/SiO$_2$ (300 nm) with predefined wells of different shapes created by reactive ion etching (RIE) on the SiO$_2$. Figure S1 shows a SEM image of the pattern of the holes. The diameter of the circular wells are between 0.5 and 3 microns and the depth is 300 nm, as the SiO$_2$ thickness.

Drumheads were prepared by deterministic all-dry transfer with a PDMS stamp, as described in references [1, 2]. There are two main advantages of using this preparation method with respect to standard microexfoliation [3] for this experiment: the first one is its higher yield covering circular wells, and the second one is that 2D material layers stamped over the holes present a very uniform and low pretension.

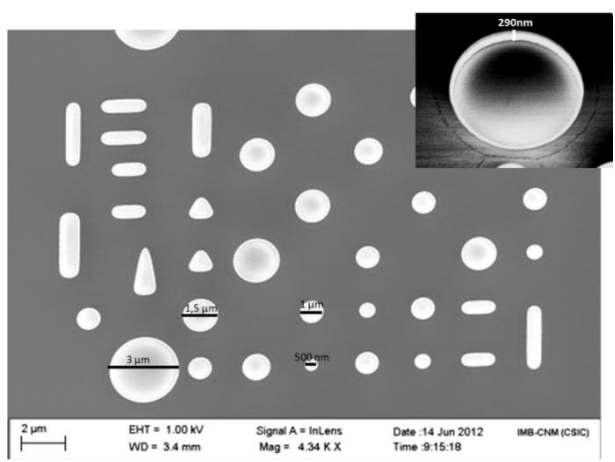

**Figure S1.** SEM image of the substrate after all the processes and a zoom in one of the holes to appreciate the verticality of the walls.

## SI2. AFM TIPS

In order to have a constant and well defined contact geometry, we use commercial tips from NanoScience Instruments with hemispherical geometry and low wear coating of tungsten carbide with nominal final tip radius of 60 nm. The cantilever spring constants were calibrated by Sader method [4] and yielded values between 25-35 N/m.

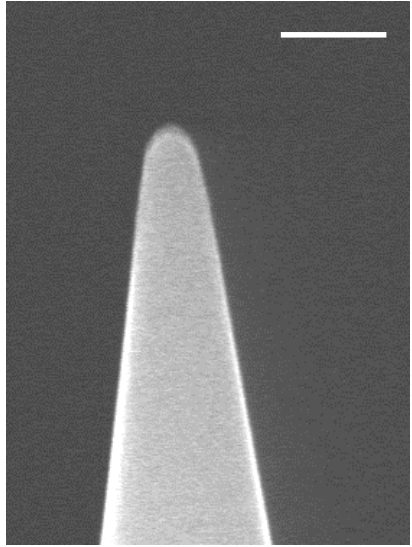

**Figure S2.** SEM image of the AFM tip used for the measurements (scale bar 200 nm).

## SI3. $F(\delta)$ CURVES

Indentation experiments were performed using a Nanotec AFM with the WSxM software [5] package. All the measurements presented here were obtained at the same loading/unloading rate of 100 nm/s. We checked that results are independent of the variation of this parameter within our experimental range.

In order to use Equation [1] in the main text, indentation has to be accurately estimated. Indentation is not a direct experimental measure; it is calculated from the differences of the relative displacement of the samples and the tip on the non-deforming $SiO_2$ substrate. This is illustrated in figure S3.

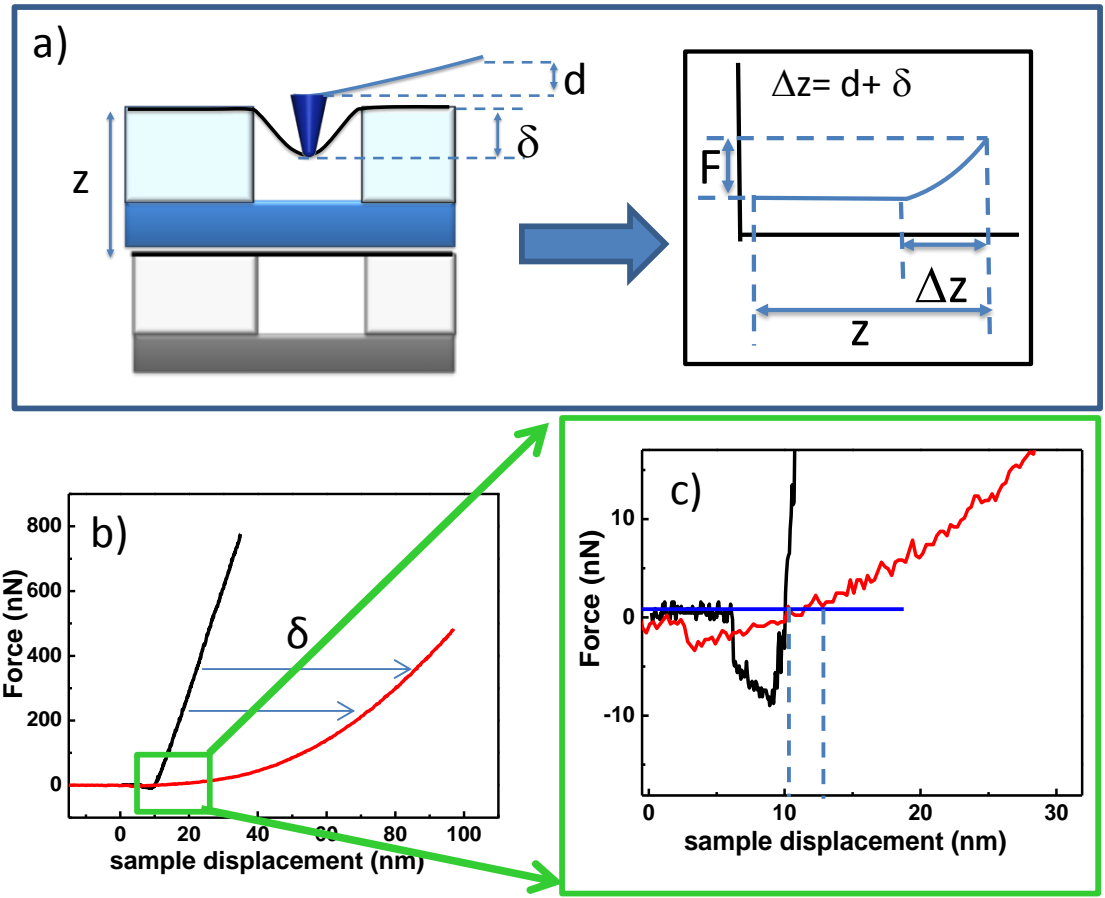

**Figure S3**. (a) Scheme illustrating the experimental procedure to extract the indentation ($\delta$) of the membrane from the displacements of the cantilever (d) and a reference substrate ($\Delta z$). (b) Force *vs.* sample displacement curve acquired on the $SiO_2$ substrate (black) and on the center of a suspended sheet (red). Panel (c) depicts a zoom of the same curve in the green rectangle. Here we can appreciate that, due to the flat character of the curves near the zero force level, we have an experimental accuracy of 2-5 nm in determining the position of zero deflection.

In order to calculate the indentation it is then critical to fix the zero displacement point, or zero force level: inaccuracy of 2-5 nm in this point leads to a 10 % error in the final calculated $E_{2D}$. According to our previous experience in analyzing indentation curves [6] the protocol followed for these curves was: we find an initial zero force level for the curve as the point where the horizontal line cuts the curve (zero cantilever deflection). A zoom in this region shows the experimental noise of the curve. To avoid noise as much as possible we find the zero by a purely instrumental fitting of the data to a high order polynomial (3$^{rd}$ order). To this end we take about 20 experimental points to the left and about 70 to right of the initial zero. The elastic modulus is then estimated according to Equation [1] in the main text.

The fitting errors for this equation for the measurements in vacuum were under 20%. In contrast for the measurements in air, the fitting error increased with time of exposure to atmospheric conditions, see figure S4. This figure also shows the decrease of the error once the sample is cleaned in vacuum through annealing (data inside a black circumference in figure S4). This suggests that the increase of the error can be caused by the layer of adsorbed water that hinders the determination of the zero indentation point of the membrane. For these cases we used a complete third order polynomial to fit the indentation curves that gives lower fitting error as it can be expected. The elastic modulus was then obtained merely from the third order term. This type of analysis had been previously used in graphene membranes very effectively [6].

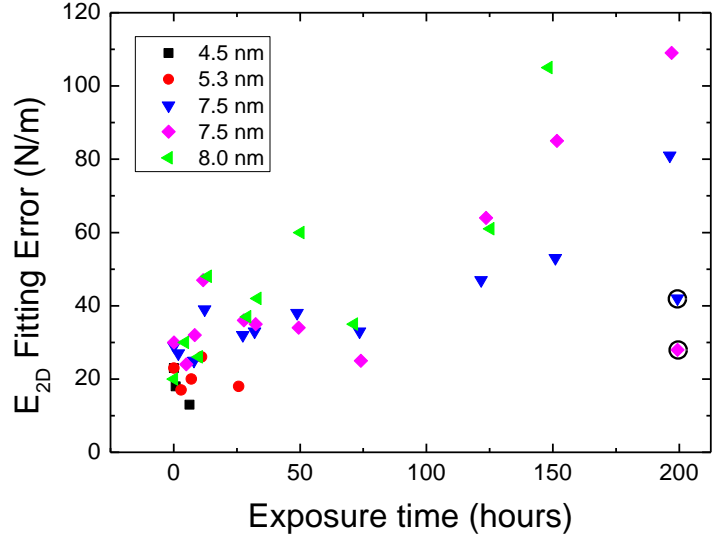

**Figure S4.** Error of the fitting of the indentation curves to a full third polynomial *vs.* the time of exposure. The data plotted at 0 hours correspond to measurements in high vacuum (before exposure), and the ones plotted inside a black circumference correspond to data acquired after the entire exposure and the subsequent heating of the sample in high vacuum.

## SI4. PRE-STRESS INFLUENCE IN $k_{flake}$ *vs.* $t^3/R^2$

Equation [2] of the main text assumes that the flakes used to calculate the elastic modulus have the same pre-tension $\sigma_0$. As we apply this expression to different flakes, we relax this condition which can result in a source of error. In order to quantify the possible error introduced by this assumption, we have numerically simulated indentations curves for 30 flakes in the low indentation regime (see figure S5(a)). The drumheads have random thicknesses between 4 and 30 nm. Each of these nanosheets has an elastic modulus of 52 GPa (value experimentally obtained by this method) and a pre-tension that can randomly vary between 0 and 0.3 N/m (the maximum pre-tension obtained from Equation [1] in the main text). As it can be seen in figure S5(b), the noise induced by the set of random pretensions have a similar dispersion to that of the experimental data. In fact, the error of $E_{3D}$ and $\sigma_0$ obtained from simulated data and experiments have the same value (±6 GPa and ±0.02 N/m for the elastic modulus and the pre-tension respectively). This suggests that the main source of noise in our data comes from the random pretensions among the different flakes.

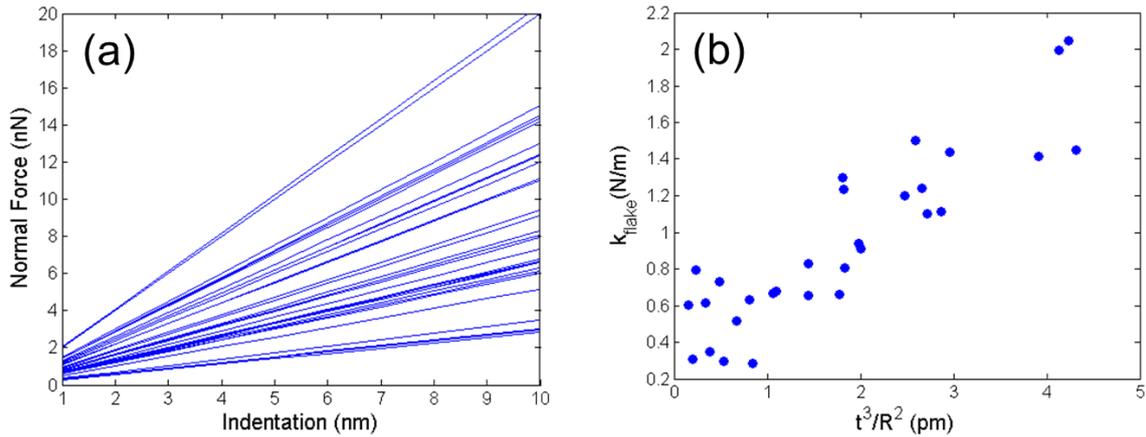

**Figure S5.** (a) Simulated force *vs.* indentation plots in the low indentation regime for flakes with growing thickness. The lines are spanned by random pre-tensions. (b) Plot of the stiffness of the flakes, calculated as the slopes of the lines in figure (a), *vs.* $t^3/R^2$. For constant pre-tension this plot should be a perfect straight line. The noise reflects the contribution of the random pre-tensions, which is similar to our experimental noise.

## SI5. AFM TOPOGRAPHIC IMAGES OF FLAKES: IN HIGH VACUUM, AIR-EXPOSED AND ANNEALED IN HIGH VACUUM

During the exposure of few-layer BP flakes to atmospheric conditions, oxidized phosphorus species and water adsorption occur. The latter (at least) seems to be somehow dependent on the flake thickness. This is shown in figure S6, where the flake show different topographic features depending on its thickness. Before exposure, the thickness of the flake was around 6 nm in the area covered by blobs while the flat part of the flake was 4.5 nm thick. Therefore, during exposure to air, the measure of the real thickness of BP flakes by AFM becomes infeasible.

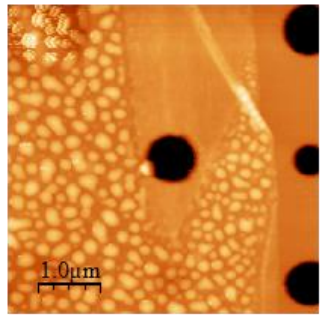

**Figure S6.** AFM topography of a flake after being exposed 120 hours to atmosphere (temperature around 22°C and average relative humidity of 47 %). Previous to the exposure, the flake was imaged by AFM showing a thickness of 6 nm in the area covered by blobs while the rest of the flake had a thickness of 4.5 nm. The drumhead was spontaneously broken during the exposure to air.

In order to measure by AFM the real thickness of the flakes after the exposure to ambient conditions, the samples were annealed in high vacuum. This allows the removal of adsorbed water. The total time of exposure to ambient conditions was nearly 200 hours. Annealing of the samples was performed in high vacuum at 230ºC during 15 hours. Afterwards, the topography of the samples were measured by AFM at room temperature maintaining the vacuum environment. Figure S7 shows the topography AFM images and profiles of a flake before the exposure and after exposure and pumping/heating. According to the profiles, the thickness of the flake has not experienced noticeable changes, while the topographic images show the appearance of some left-over residues.

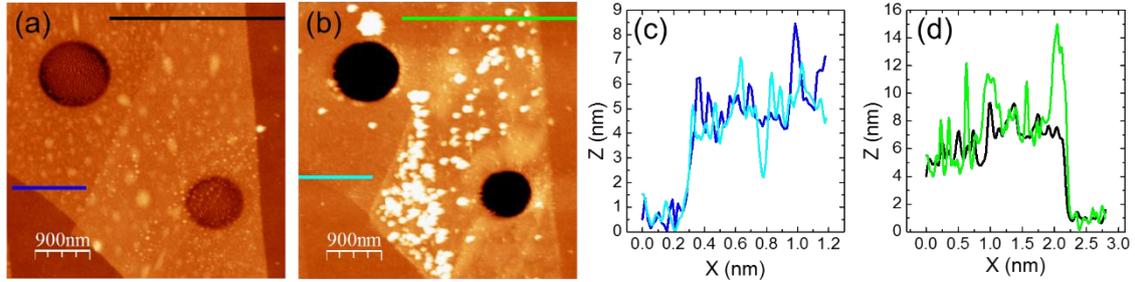

**Figure S7.** Topographic changes in BP flakes after nearly 200 hours of exposure to ambient conditions and annealing at high vacuum. (a) Topography AFM image (the same image as figure 1(b) in the main text) of a BP flake in high vacuum, before the exposure to atmosphere. (b) Topography AFM image of the same flake shown in (a), acquired also in high vacuum but once the sample has been in contact to air for almost 200 hours, and subquently heated in high vacuum. Both membranes appear broken due to indentations experiments performed until rupture in high vacuum before exposure to air. (c) Topographic profiles along the blue and cyan lines in (a) and (b). (d) Topographic profiles along the black and green lines in (a) and (b).

## SI6. PASSIVATION MODEL

For the purpose of understanding the different environmental effects in the values obtained for $E_{3D}$ in thin (< 6 nm) and thick (> 7 nm) drumheads, we propose an elementary model of the passivation process. It is based in the following assumptions (see figure S8(a)):

(i) The total thickness of every BP flake during the exposure time is constant and equal to the initial thickness of the BP flake measured in high vacuum before the passivation process starts ($t_{BP}^{ini}$):

$$t_{BP}^{ini} = t_{pass}(\tau) + t_{BP}(\tau) = const.$$

where $\tau$ is the exposure time, $t_{pass}$ the thickness of the outer passivated BP and $t_{BP}$ the thickness of the underlying pristine BP flake. This assumption is supported by the experimental preservation of flakes' height shown in figure S7.

(ii) The growth of the passivation layer takes place at a speed that exponentially decays with time, in other words, the passivation process experiences saturation, having a maximum passivation depth, $t_{pass}^{max}$. This is suggested by the data of thicker flakes at advanced exposure in figure 4 in the main text. This leads to the following relations:

$$t_{pass}(\tau) = t_{pass}^{max} \cdot [1 - exp(-\tau/\tau_c)]$$

$$t_{BP}(\tau) = t_{BP}^{ini} - t_{pass}(\tau)$$

where $\tau_c$ is a passivation characteristic time.

(iii) The elastic moduli of the pristine and passivated BP are constant with time and equal to $E_{3D}^{BP}$ and $E_{3D}^{pass}$ respectively, being $E_{3D}^{pass} = \alpha \cdot E_{3D}^{BP}$, $\alpha < 1$. According to [7] a value of $\alpha = 0.66$ is expected for phosphorene, considering the average of the elastic modulus of both in-plane directions at two stages: when the oxidation ratio is 0% and 100%.

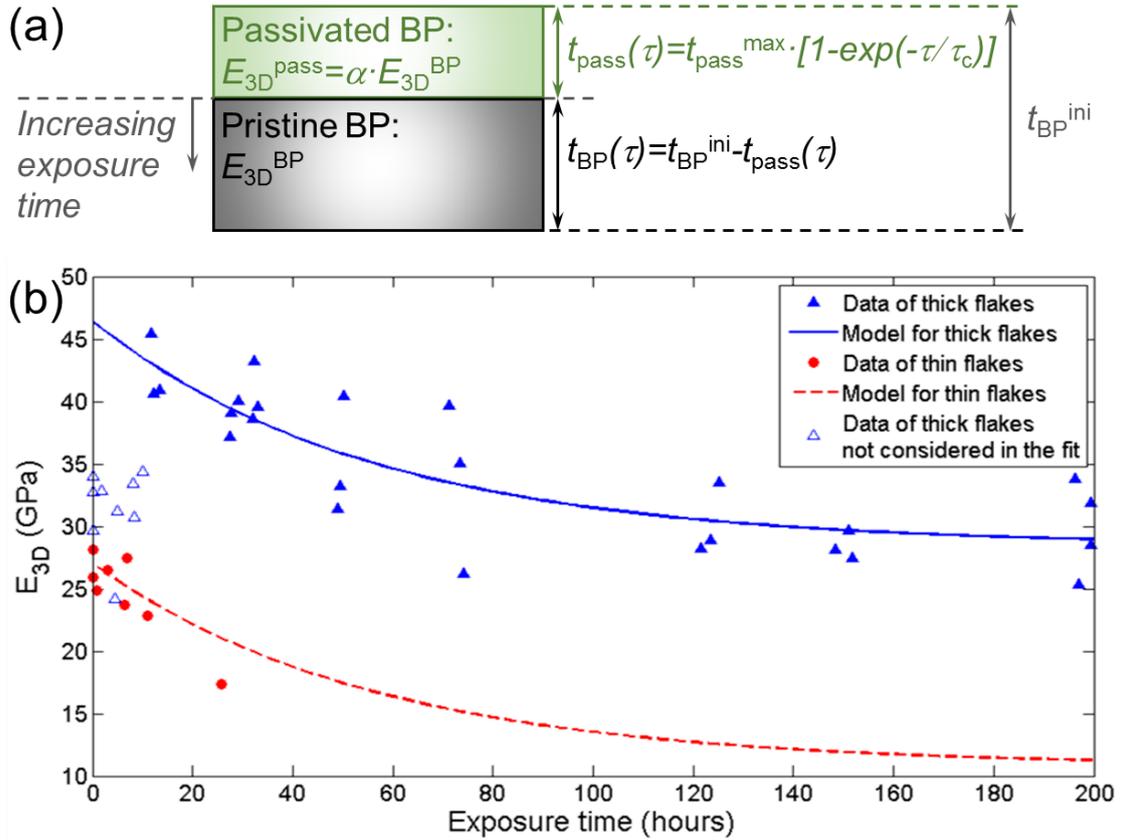

**Figure S8.** (a) Diagram of the passivation model proposed herein. (b) Temporal evolution of the elastic modulus of few-layer BP flakes under exposure to ambient conditions. The symbols correspond to experimental data (blue solid and empty triangles for drumheads thicker than 7 nm, and red circles for drumheads thinner than 6 nm). Blue line corresponds to the fit of the blue solid triangles to the proposed model and the dashed red line depicts the model for the thinner flakes.

Considering the mentioned assumptions, the fact that the $E_{2D}$ of a layered material is the sum of the $E_{2D}$ of each layer ($E_{2D}=E_{2D}^{BP}+E_{2D}^{pass}$), and that the magnitude that is obtained from our measurements is $E_{2D}=E_{3D}\cdot t = E_{3D}^{BP}\cdot t_{BP}(\tau)+E_{3D}^{pass}\cdot t_{pass}(\tau)$, the elastic modulus measured during the exposure time is:

$$E_{3D}^{measured}(\tau) = \frac{E_{3D}^{BP}}{t_{BP}^{ini}}\{t_{BP}^{ini} - t_{pass}^{max}(1-\alpha)[1 - exp(-\tau/\tau_c)]\}$$

The initial increase in the elastic modulus of the thicker flakes in figure 4 of the main manuscript is in agreement with the evolution of $E_{3D}$ of phosphorene under oxidation process according to Hao *et al.* [7]. In this reference an increase of $E_{3D}$ in the armchair direction until an oxidation ratio of 25% is ascribed to small relaxations of phosphorene that occur due to chemisorbed oxygen atoms. This increase is not observed in our data of the thinner flakes, probably because it took place in the short exposure prior to the measurements in high vacuum.

Figure S8(b) shows the evolution of the elastic modulus of exposed drumheads with the exposure time. There are two sets of data. One corresponds to the flakes thicker than 7 nm of figure 4, whose average thickness is 7.6 nm, plotted with blue empty and solid triangles. The second set comprises the data of the flakes thinner than 6 nm of figure 4, with an average thickness of 4.9 nm, plotted with red dots. In order to fit the data to the suggested model, firstly, only the data of the thick drumheads are considered, since they present a longer evolution. The data acquired in the first 10 hours (empty triangles) are discarded to avoid the aforementioned increase of $E_{3D}$ at low oxidation rates. Hence, the solid triangles are fitted to the exponential decay expression of $E_{3D}^{measured}(\tau)$ given above. The following values are fixed: $t_{BP}^{ini} = 7.6$ nm and $\alpha = 0.66$. This fitting, depicted with a blue line in figure S8(b), yields $E_{3D}^{BP} = 46\pm5$ GPa, $t_{pass}^{max}=9\pm5$ nm and $\tau_c=60\pm20$ hours. The value for elastic modulus of pristine BP coincides with the one given in the paper in vacuum environment. For the thinner flakes, we use the obtained values for $t_{pass}^{max}$ and $\tau_c$ and $t_{BP}^{ini} = 4.9$ nm, $E_{3D}^{BP} = 27$ GPa (provided by the intercept of the linear fit of the first seven data) and $\alpha = 0.66$ again. In this case the red dashed line in figure S8(b) is obtained.

The proposed model and the characteristic values obtained from it, result in some important consequences. Firstly, all the flakes considered in this section, after a long enough exposure, would be completely oxidized, considering that their thicknesses are lower than the maximum passivation depth. Secondly, and more important, every few-layer BP nanosheet thicker than 9 nm would undergo a passivation process until the passivated layer reaches that thickness, similarly to the natural passivation of some other materials such as stainless steel, aluminum, cooper, etc. However, this expected behaviour could deviate from some other actual tendencies due to the dispersion of our data. In fact, this model would not apply for very thick BP flakes (around 100 nm or more), in which water ends covering the BP surface causing a layer-by-layer etching as was reported in [8] and confirmed by us through some optical images of flakes thicker than 100 nm. Moreover, the presented model does not consider the likely circumstance of having passivation on the surface of the membrane facing down, nor simultaneously passivation and etching processes. This later situation was not reflected by our experimental results in few-layer BP flakes.